\documentclass[12pt]{article}
\usepackage{lineno}
\linenumbers
\usepackage{epsfig}
\usepackage{subfigure}
\begin{document}

{\LARGE \begin{center}Aluminum as a source of background in low background experiments \end{center}}
\begin{center}
B.\,Majorovits$^{a,}$\footnote{ corresponding author. Tel.: +49 89 32354 262, FAX: +49 89 32354 528. Adress: MPI f\"ur Physik, F\"ohringer Ring 6, 80805 M\"unchen, E-mail: bela@mppmu.mpg.de}, I.\,Abt$^a$, M.\,Laubenstein$^b$, O.\,Volynets$^a$
\end{center}

\centerline{\it\small $^a$ MPI f\"ur Physik, F\"ohringer Ring 6, 
80805 Munich, Germany}

\centerline{\it\small$^b$ Laboratori Nazionali del Gran Sasso, INFN, S.S.17/bis, km 18+910, I-67100
Assergi (AQ), Italy}

\begin{abstract}
Neutrinoless double beta decay would be a key to understanding the nature of neutrino masses. 
The next generation of High Purity Germanium experiments will have to be operated with a background rate of better than 10$^{-5}$\,counts/(kg\,y\,keV) in the region of interest around the Q value of the decay.
Therefore, so far irrelevant sources of background have to be considered.
The metalization of the surface of germanium detectors is in general done with 
aluminum.
The background from the decays of $^{22}$Na, $^{26}$Al, $^{226}$Ra and $^{228}$Th introduced by this metalization is discussed. 
It is shown that only a special selection of aluminum can 
keep these background contributions acceptable.

\end{abstract}

\newpage
\section{Introduction}
High Purity Germanium (HPGe) detectors are extremely sensitive spectrometers, used to search for rare events, such as neutrinoless double beta decays or dark matter scattering off nuclei \cite{cite:gerda, cite:cdms, cite:edelweiss}.
In order to rule out the possibility of an inverted neutrino mass hierarchy, a neutrino with a Majorana mass as low as 10 meV has to be excluded.
The low event rates expected, require very low background environments and a very high target mass. 
 
An HPGe based neutrinoless double beta decay experiment would need an active mass of $\approx$ 1\,ton and a background level below
10$^{-5}$\,counts/(kg\,y\,keV) around the Q-value of the double beta decay of $^{76}$Ge at 2039 keV.
This requires a reduction of the background rate by three to four orders of magnitude with respect to the current state of the art.
This necessitates the investigation of so far unimportant background contributions.
Here, the effects of $^{26}$Al and $^{22}$Na, cosmogenically produced in aluminum, are investigated.

As it is not trivial to obtain 
radio-pure aluminum \cite{cite:heusser}, it is usually avoided close to the active elements of low background experiments.
However, aluminum is commonly used to metallize HPGe detector surfaces. For low background applications, great care is often taken to use especially selected "Ultra Low Background" (ULB) aluminum \cite{cite:loaiza, cite:koehler}.
However, even if U- and Th- free aluminum is used, $^{26}$Al, cosmogenically produced, is not removed during the refinement process.

The isotope $^{26}$Al has a Q-value of 4.0~MeV and a half life of 7.17\,$\cdot 10^5$\,years \cite{cite:aluminum_half_life}. In 81.7 \%, it decays by emission of a positron, an electron and a 1808.7~keV gamma to $^{26}$Mg. Thus, it can deposit around $\approx$ 2~MeV inside the detector.

The isotope $^{22}$Na can also be cosmogenically produced in aluminum.
It has a Q-value of 2.84~MeV and a half life of 2.6 years \cite{cite:aluminum_half_life}. In 90.3 \%, it decays by emission of a positron and a 1274.5~keV gamma to $^{22}$Ne. 
During the refining process leading to commercially available aluminum, it is efficiently separated. However, as the half life is relatively short, equilibrium between production and decay rate is regenerated within the relatively short time span of a few years. 

\section{Cosmogenic production of $^{26}$Al and  $^{22}$Na from  $^{27}$Al}

Aluminum consists $\approx$\,100\,\% of the isotope $^{27}$Al. 
Small quantities of the ground state of $^{26}$Al, $^{26}$Al-g, can be produced cosmogenically through the reactions $^{27}$Al(n,2n)$^{26}$Al-g and $^{27}$Al(p,np)$^{26}$Al-g.
The isotope $^{22}$Na can be produced by the reactions $^{27}$Al(n,2p4n)$^{22}$Na and $^{27}$Al(p,3p3n)$^{22}$Na.

If the energy dependent production cross-sections, $\sigma(E)$, for the reactions under consideration and the differential neutron and proton fluxes as a function of energy, $\frac{\partial \Phi _{n/p}}{\partial E}$, are known,
the total production rates can be calculated as

\begin{equation}
dN/dt = N_0 \int _0^{\infty} \frac{\partial \Phi _{n/p}}{\partial E} \cdot \sigma(E)  dE ,
\end{equation}

where N$_0$ is the number of target nuclei.
In equilibrium, the decay and production rates of the considered isotope are equal.

Additionally, production of  $^{26}$Al and $^{22}$Na through negative muon capture and spallation reactions induced by fast through-going muons is possible.

\subsection{Production cross-sections}
\begin{figure}
\begin{center}
\epsfig{file=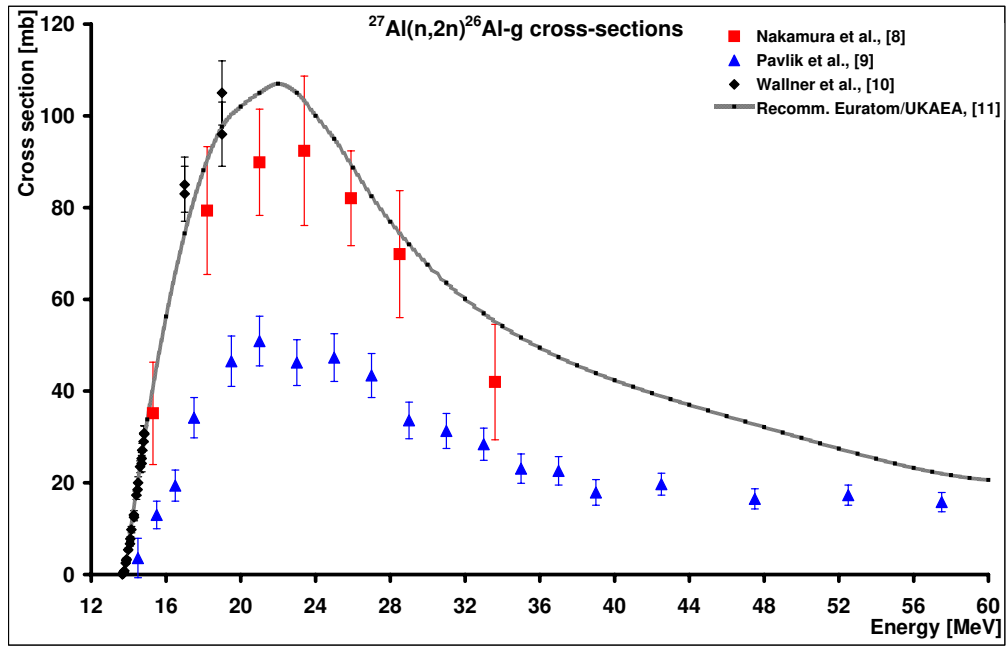, height=12cm}
\end{center}
\caption{Measured production cross-sections for the reaction $^{27}$Al(n,2n)$^{26}$Al-g \cite{cite:nakamura, cite:pavlik, cite:wallner}. The vertical bars represent the  1\,$\sigma$ statistical uncertainties. If the bars are not visible, they are smaller than the symbols. The solid line represents the recommendation from \cite{cite:ukaea} used  to calculate the production rate. \label{fig:26al_prod_cross_sections_n}}
\end{figure}

The measured production cross-sections of the $^{27}$Al(n,2n)$^{26}$Al-g reaction above the production threshold of 13.55\,MeV up to neutron energies of 60\,MeV are summarized in Fig.\,\ref{fig:26al_prod_cross_sections_n}.
Above 15\,MeV, three measurements exist \cite{cite:nakamura, cite:pavlik,cite:wallner}. 
The cross-section peaks at 20\,MeV according to Nakamura et  al. \cite{cite:nakamura} and Pavlik et al. \cite{cite:pavlik}. In Wallner et al. \cite{cite:wallner} the highest available neutron energy was 19\,MeV. The measured cross-sections of the latter reference are consistent with the data shown in Nakamura et al. They are roughly a factor of two higher than the 
cross-sections shown in Pavlik et al. 
For the calculation of the number of cosmogenically produced $^{26}$Al nuclei by neutrons, a polynomial fit to the cross-sections from Wallner et al. and Nakamura et al. is used up to 20~MeV and values recommended by Euratom/UKAEA \cite{cite:ukaea} were used for energies between 20\,MeV and 60\,MeV. The recommended excitation function is also shown in Fig.\,\ref{fig:26al_prod_cross_sections_n}. Up to 20\,MeV, the recommendation coincides with the polynomial fit to the Wallner et al. and Nakamura et al. data. Above 60\,MeV, a constant cross-section of 20\,mb was assumed.

\begin{figure}
\begin{center}
\epsfig{file=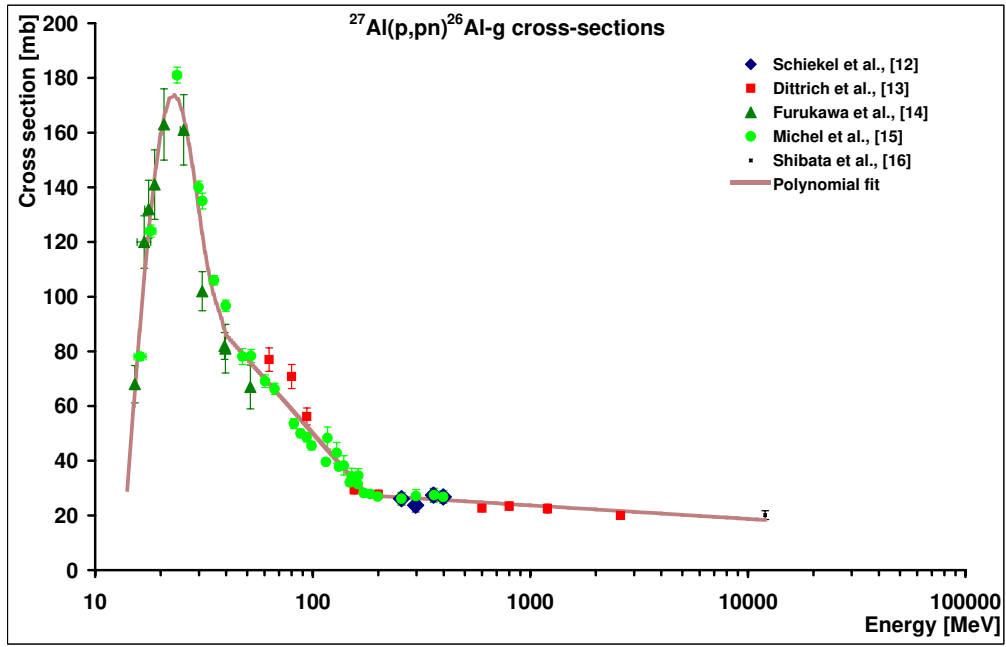, height=12cm}
\end{center}
\caption{Measured production cross-sections for the reaction $^{27}$Al(p,pn)$^{26}$Al-g \cite{cite:schiekel, cite:dittrich, cite:furukawa, cite:michel, cite:shibata}. The bars represent the 1\,$\sigma$ statistical uncertainties. If the bars are not visible, they are smaller than the symbols. The solid line represents the polynomial fit to calculate the production rate.\label{fig:26al_prod_cross_sections_p}
}
\end{figure}

Measured cross-sections \cite{cite:schiekel, cite:dittrich, cite:furukawa, cite:michel, cite:shibata} for the reaction $^{27}$Al(p,pn)$^{26}$Al-g, including a polynomial fit, are shown in Fig.\,\ref{fig:26al_prod_cross_sections_p}. The fit was used for the calculation of the production rate through this excitation channel.

\begin{figure}
\begin{center}
\epsfig{file=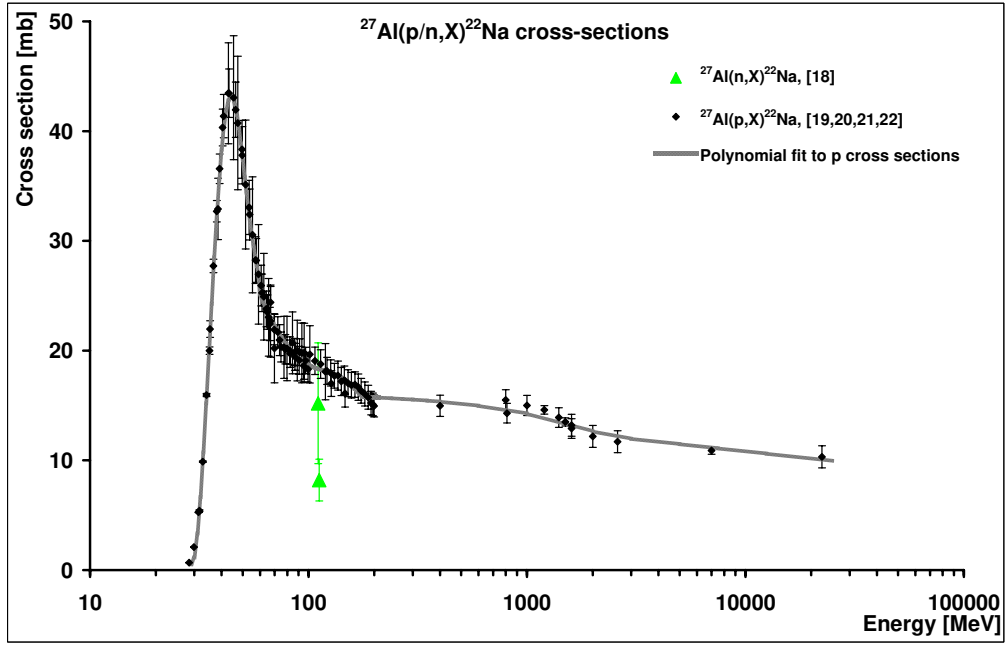, height=12cm}
\end{center}
\caption{Measured production cross-sections for the reactions 
$^{27}$Al(n,2p4n)$^{22}$Na (triangles) \cite{cite:sisterson_2} and 
$^{27}$Al(p,3p3n)$^{22}$Na (spades) \cite{cite:buthelezi, cite:titarenko, cite:morgan, cite:steyn} . The bars represent the statistical  1\,$\sigma$ uncertainties. If the bars are not visible, they are smaller than the symbols. The solid line represents the polynomial fit used to calculate the production rates.
\label{fig:22na_prod_cross_sections}
}
\end{figure}

There are little data on $^{22}$Na production cross-sections from the reaction  $^{27}$Al(n,2p4n)$^{22}$Na. 
An integrated cross-section of about 8\,mb \cite{cite:sisterson} has been measured from the threshold of 23\,MeV up to 750\,MeV. Cross-sections of (15.2$\pm$5.5)\,mb and (8.2$\pm$1.9)\,mb at energies of 110.8\,MeV and 112\,MeV, respectively, have also been reported \cite{cite:sisterson_2}. These values agree within the uncertainties. They are displayed in Fig.\,\ref{fig:22na_prod_cross_sections}.
The available data for the reaction $^{27}$Al(p,3p3n)$^{22}$Na \cite{cite:buthelezi, cite:titarenko, cite:morgan, cite:steyn} are also presented in
Fig.\,\ref{fig:22na_prod_cross_sections}. 

As no energy dependent measurements of the cross-section for the reaction $^{27}$Al(n,2p4n)$^{22}$Na were available, it was assumed that the cross-sections for neutrons and protons as a function of energy are the same. For energies above the Coulomb barrier of a reaction, this assumption is justified due to isospin symmetry.
The Coulomb barrier is around a few MeV for the reactions under investigation.
For the calculations of the production rates from the reaction $^{27}$Al(n,2p4n)$^{22}$Na and $^{27}$Al(p,3p3n)$^{22}$Na, a polynomial fit to the proton excitation function data, displayed in Fig.\,\ref{fig:22na_prod_cross_sections}, was used.

\subsection{Secondary neutron and proton fluxes at sea level}

\begin{figure}
\begin{center}
\epsfig{file=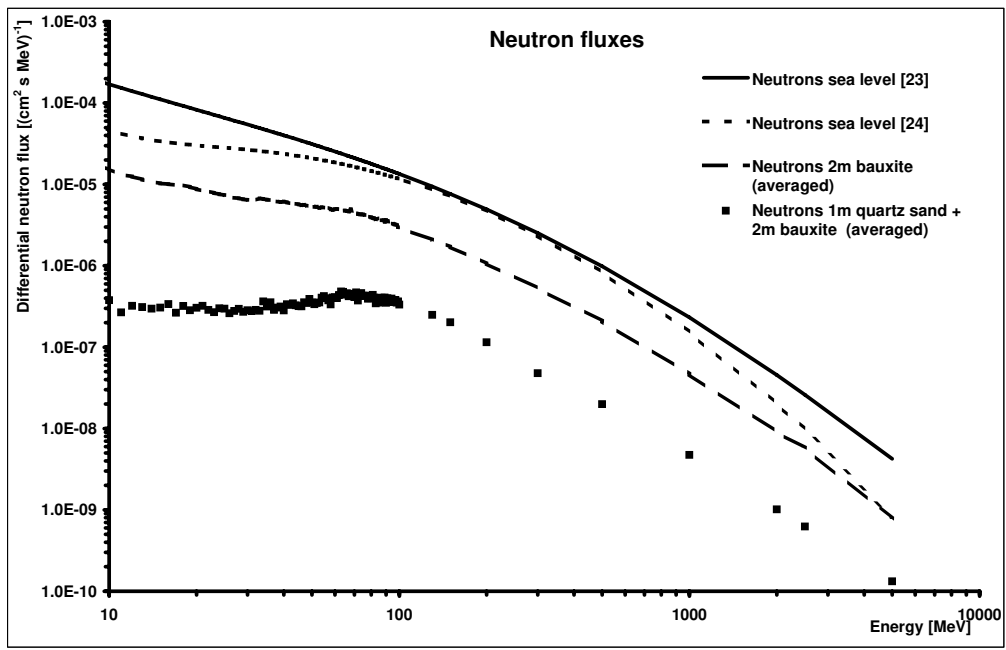, height=12cm}
\end{center}
\caption{Differential parameterized neutron fluxes around New York, USA \cite{cite:ziegler,cite:neutrons_gordon} and the simulated neutron spectra averaged over a 2\,m thick bauxite layer with and without 1\,m of quartz sand as top soil.  \label{fig:neutron_fluxes}
}
\end{figure}

Two sets of neutron flux measurements \cite{cite:ziegler, cite:neutrons_gordon} at sea level around New York are shown in Fig.\,\ref{fig:neutron_fluxes}.
The measurements agree within an order of magnitude. However, for energies below 100\,MeV and above 1000\,MeV, they still differ significantly.

\begin{figure}
\begin{center}
\epsfig{file=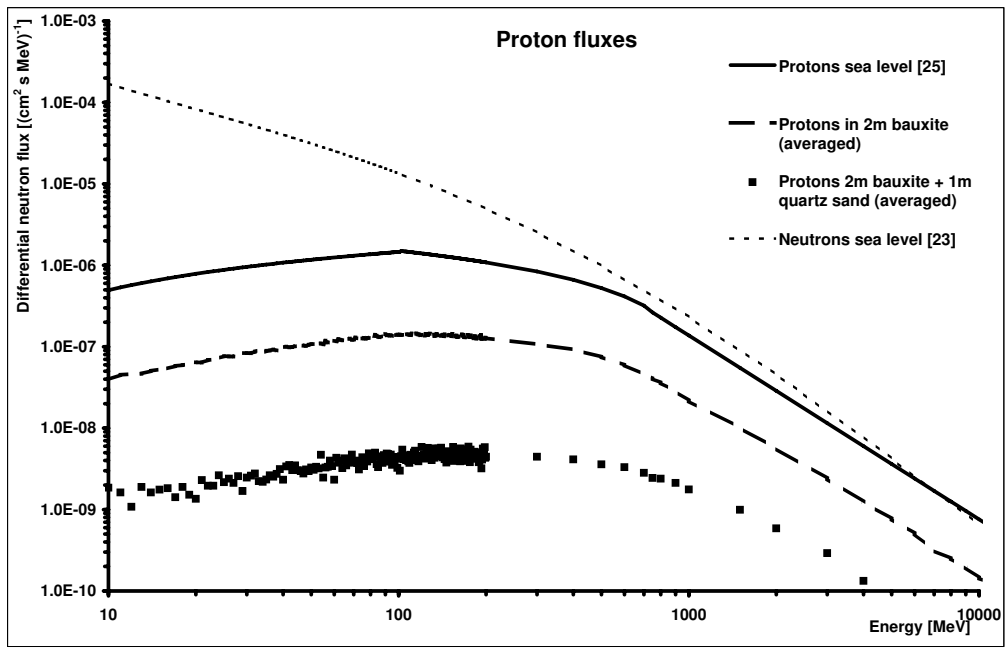, height=12cm}
\end{center}
\caption{Differential parameterized proton flux \cite{cite:filthuth} and the simulated proton spectra averaged over a 2\,m thick bauxite layer with and without 1\,m of quartz sand as top soil. The neutron flux at sea level \cite{cite:ziegler} is given as reference.
\label{fig:proton_fluxes}
}
\end{figure}

The secondary proton flux at sea level is significantly lower than the neutron flux at energies relevant for the production of $^{26}$Al and $^{22}$Na. A polynomial fit to the proton spectrum measured in Karlsruhe, Germany \cite{cite:filthuth} is shown in Fig.\,\ref{fig:proton_fluxes}. 

For the calculation of the expected production rates, the more conservative neutron spectrum from \cite{cite:ziegler} and the fit to the proton spectrum were used, unless otherwise stated.

\subsection{Exposure of bauxite to secondary cosmic rays}
Aluminum is refined from bauxite deposits that were 
predominantly formed by weathering \cite{cite:valeton}
during several periods in Earth history.
The currently mined deposits consist mainly of lateral sheets
that have rested on or close to the surface for the past million years.
The top-soil overburden
above the bauxite layer is often negligible \cite{cite:bauxite_composition} or has a thickness of less 
than a meter for major mining sites \cite{cite:bauxite}.
The bauxite layers themselves are usually as thin as 2\,m to 4\,m \cite{cite:bauxite}.
Underground mines, exploiting pockets of bauxite, are mainly
located in Southern Europe and Hungary and have a small market share.

The top soil above the bauxite layers of mining sites and the self 
shielding of the 2 -- 4\,m thick bauxite layers reduce the secondary 
cosmic ray flux and thus the cosmogenic activation of the aluminum yielding $^{26}$Al and $^{22}$Na.
In order to account for the shielding by the top soil and the self shielding of 
the bauxite layer, a Monte Carlo simulation was performed. 
The MaGe \cite{cite:MaGe} framework based on GEANT4 \cite{cite:geant4} 
was used.

Two scenarios, typical for major bauxite mining sites, were taken into 
account: a top soil free mining site (like WEIPA, Australia \cite{cite:bauxite_composition}) and a site with 1\,m quartz sand overburden. The composition of the simulated bauxite was chosen according 
to the one reported for the WEIPA site \cite{cite:bauxite_composition}.

For both scenarios, top soil free and 1\,m top soil, the neutron and proton fluxes were simulated at different depths in the bauxite layer: 
for depths between 0\,m and 0.5\,m in steps of 0.1\,m and for 
depths between 0.5\,m and 2.0\,m in steps of 0.5\,m. 
The fluxes calculated for the different depths were averaged to obtain 
the mean neutron spectra for a bauxite layer with 2\,m thickness with and without top soil.

The spectra for neutrons and protons are shown in 
Fig.\,\ref{fig:neutron_fluxes} and Fig.\,\ref{fig:proton_fluxes}, 
respectively. The averaged spectra were used for calculating 
realistic $^{26}$Al and $^{22}$Na production rates within a 2\,m 
bauxite layer. 

\subsection{Calculated $^{26}$Al and $^{22}$Na production rates}
\begin{figure}
\begin{center}
\epsfig{file=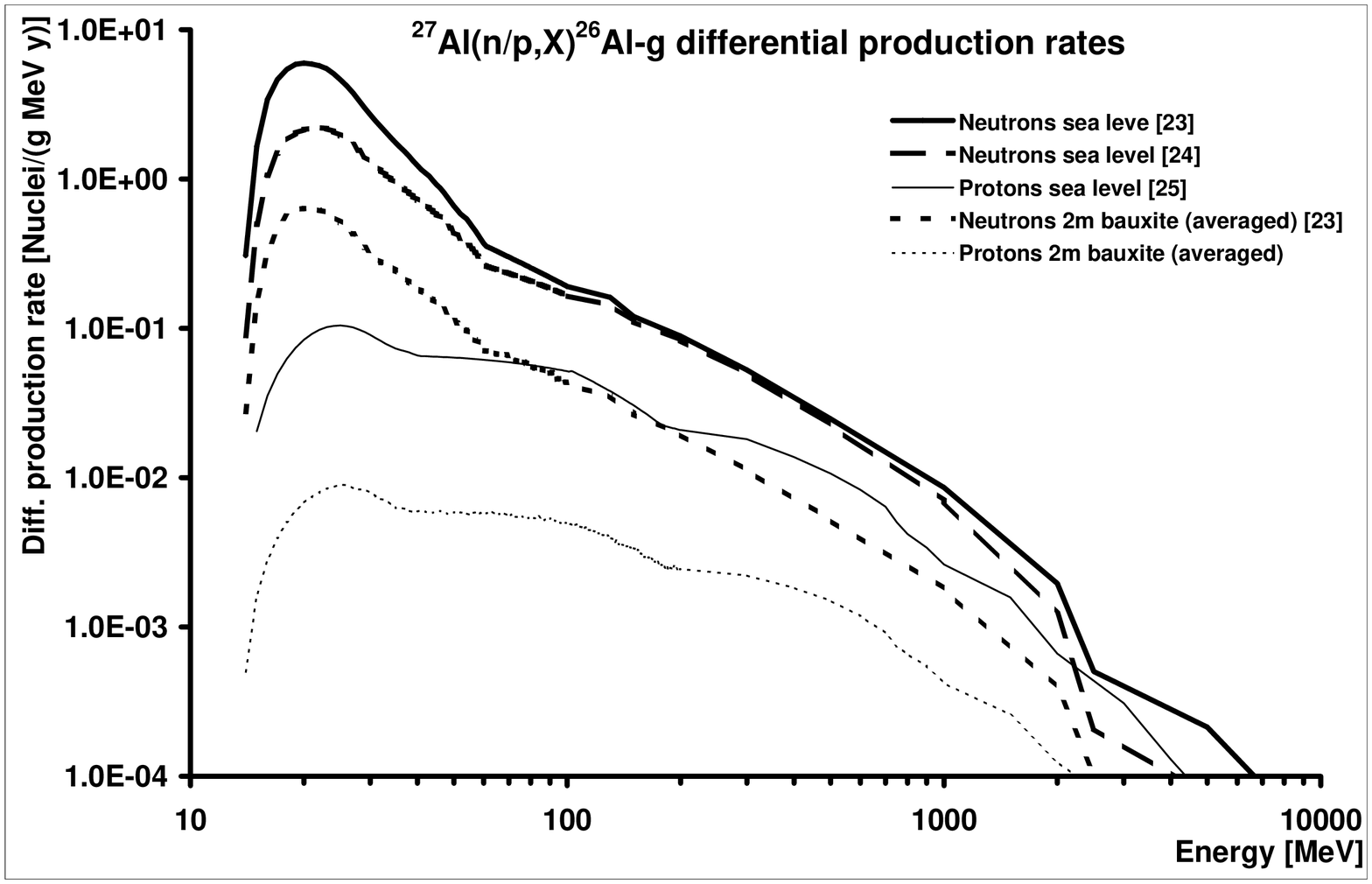, height=12cm}
\end{center}
\caption{Calculated differential  production rates of $^{26}$Al in $^{27}$Al due to secondary cosmic protons and neutrons at sea level  and averaged over a 2\,m thick bauxite layer without top soil. \label{fig:26alprod_rates_diff}
}
\end{figure}

\begin{figure}
\begin{center}
\epsfig{file=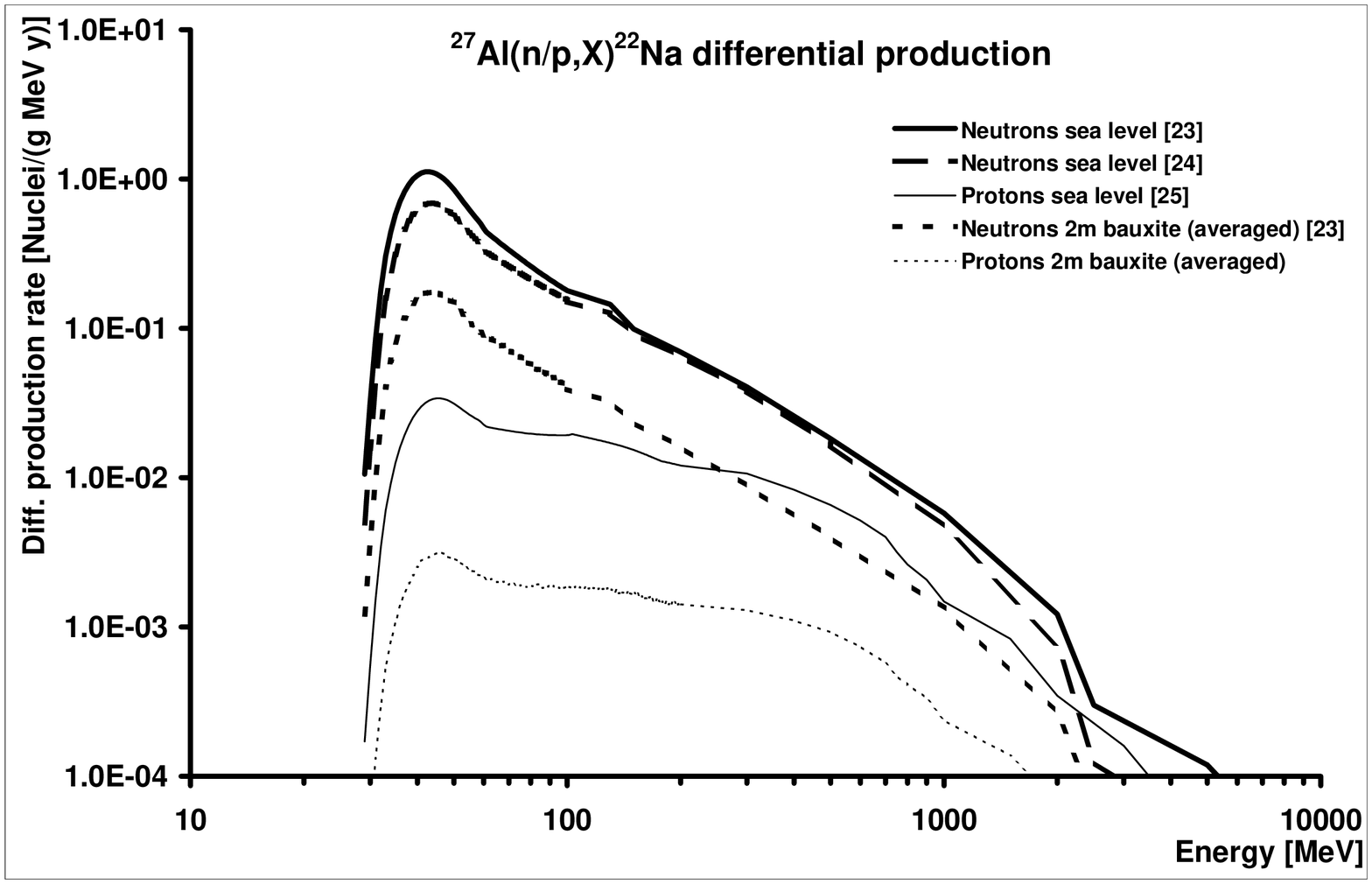, height=12cm}
\end{center}
\caption{Calculated differential production rates of $^{22}$Na in $^{27}$Al due to secondary cosmic protons and neutrons at sea level and averaged over a 2\,m thick bauxite layer without top soil. \label{fig:22naprod_rates_int}
}
\end{figure}

The calculated differential 
production rates of $^{26}$Al and $^{22}$Na for the previously described neutron and proton
spectra at sea level and for different shielding scenarios are shown in Fig.\,\ref{fig:26alprod_rates_diff} and Fig.\,\ref{fig:22naprod_rates_int}, respectively.
The production rate is significantly lower for the spectrum of \cite{cite:neutrons_gordon} than for the selected standard \cite{cite:ziegler} below 100\,MeV. 

Self shielding by a 2\,m bauxite layer does reduce the differential production rates of  $^{26}$Al and $^{22}$Na by a factor of roughly 4\,--\,10 with respect to 
the sea level production rate, depending on the neutron and proton energy.  

The integrated production rates of $^{26}$Al and $^{22}$Na are given in Table \ref{tab:prod_rates}.
The reference neutron flux \cite{cite:ziegler} at sea level without overburden results in a production rate of 142 (56) $^{26}$Al ($^{22}$Na) nuclei per year and gram aluminum. The reduced neutron flux \cite{cite:neutrons_gordon} at sea level results in a production rate of roughly 80 (43) $^{26}$Al ($^{22}$Na) nuclei per gram aluminum per year.
The higher production threshold of $^{22}$Na reduces this difference in the production rate of $^{22}$Na compared to the one of $^{26}$Al. 

The mean $^{26}$Al ($^{22}$Na) production rate due to the \cite{cite:ziegler} in a 2\,m thick bauxite layer is 21 (11) $^{26}$Al ($^{22}$Na) nuclei per year and gram aluminum.
 

If another 1\,m of quartz sand overburden is assumed, the production rate through secondary neutrons drastically drops to 1.4 (1.0) $^{26}$Al ($^{22}$Na) nuclei per year and gram of aluminum.

The expected production rate at sea level without overburden from protons is 17 (3) $^{26}$Al ($^{22}$Na) nuclei per gram aluminum per year. It becomes negligible if self shielding or an additional layer of top soil are taken into account.


In total in equilibrium in 
a 2\,m bauxite layer without top soil 23 $^{26}$Al decays 
per gram aluminum 
per year are expected. At sea 
in equilibrium 65 $^{22}$Na decays are expected per gram aluminum per year.

Muon capture and spallation reactions induced by through-going muons can also lead to the production of $^{26}$Al and $^{22}$Na in quartz. The contribution due to this production channel at sea level is roughly 4\% \cite{cite:heisinger}. As neutrons and protons are absorbed more efficiently in matter than muons, the fraction of $^{26}$Al and $^{22}$Na produced by negative muon capture and fast muons increases with increasing overburden.

To estimate the realistic contribution of the production rate due to muons, 
the production rates deeper inside the bauxite layer have to be considered.
This was done for the case of no top soil and a 2\,m thick bauxite layer. The density of bauxite is very sensitive to the grain size \cite{cite:bauxite_density}. 
It ranges between 1.3\,g/cm$^3$ for fine grain up to 3.5\,g/cm$^3$ for 
dense bauxite with a large quartz content. Thus, a 2\,m bauxite layer
 will correspond to 2.6\,--7\,meter water equivalent, mwe, depending on the mining site.
At 2.6\, mwe (7\,mwe) the contribution of muons to the $^{26}$Al production 
in quartz is 10\%\,(75\%) \cite{cite:heisinger}. For deep bauxite layers with non negligible top soil, the production channel through muons will be dominating.
For the case of shallow, low density bauxite mines, considered here, the production rate through muons is small.
As there are already large uncertainties in the secondary neutron and proton fluxes, the muon contribution is neglected from now on.

\section{Comparison to measurements of cosmogenically produced $^{26}$Al in quartz}
The isotope $^{26}$Al is a well known tracer for erosion rates in geology \cite{cite:heisinger}. The production rates of $^{26}$Al have mainly been measured in quartz samples.  In glacially exposed surface layers, an average production rate of 374$\pm$28 $^{26}$Al nuclei per g quartz per year has been measured at an elevation of 3340\,m \cite{cite:al_in_Si_glacier}. This can be translated into a production rate at sea level of 36.8$\pm$2.7 (stat.) $^{26}$Al nuclei per g quartz per year. This is in good agreement with recent measurements from \cite{cite:al_in_Si_II}.
These measurements can be used to validate the calculation method used for production rates in Bauxite.

The excitation functions for $^{26}$Al in $^{28}$Si for neutrons are only 
available  for energies between 25\,MeV and 36\,MeV \cite{cite:imamura}. At higher energies, only proton excitation functions are available \cite{cite:schiekel, cite:michel_si}.
As the measured neutron \cite{cite:imamura} and proton \cite{cite:furukawa} 
cross-sections are in very good agreement at energies between 25\,MeV and 35\,MeV , it can be assumed that the neutron excitation function is about equal to the measured proton excitation function at all energies. 
At high energies, excitation functions for neutrons and protons are in general  expected to be the same due to isospin symmetry. 

Using a polynomial fit to the excitation function \cite{cite:imamura} and the unshielded neutron (proton) spectrum at sea level from \cite{cite:ziegler} (\cite{cite:filthuth}), a $^{26}$Al production rate in quartz at sea level of 44 (6) nuclei per gram quartz is expected.
Considering the large uncertainties in the secondary cosmic ray flux variations with geographic location and the uncertainties in the exposure histories of the measured samples, this is in good agreement with the experimental values.

The good agreement between calculation and measurement shows that the method to estimate the activities of $^{26}$Al and $^{22}$Na in bauxite samples can be trusted to roughly a factor of two.

\section{Background expectations due to  $^{26}$Al and  $^{22}$Na}
The contamination of any sample of aluminum depends on its history.
Equilibrium between production rate and decay of an isotope will be reached after a few half lives. This is the case after a few million (ten) years for $^{26}$Al ($^{22}$Na). 
For the case of a 2\,m bauxite layer without top soil, in equilibrium  a   $^{26}$Al activity of 23 decays per year is expected per gram aluminum, see Table \ref{tab:prod_rates}.


The contamination with $^{26}$Al is not affected by the refining process.
The expected equilibrium activity for a 2\,m bauxite deposit without 
top soil is 0.8\,mBq/kg. For ease of scaling, an activity of 
1.0\,mBq/kg was used for the background estimates.
Due to its long half life, the background from $^{26}$Al is constant during 
any experiment.

The contamination with $^{22}$Na is eliminated effectively during the refining
process. Thus for $^{22}$Na it is the exposure history after aluminum refinement that determines its activity.
After one half life (2.6 years) of exposure to cosmic rays, the decay rate is equal to half of the production rate. For no overburden at sea level, this corresponds to 1.0\,mBq/kg. This value was assumed to estimate backgrounds.
For underground experiments, the $^{22}$Na activity decreases significantly after a few years.   

Significant $^{228}$Th contaminations of the order of mBq/kg 
have been measured in ULB aluminum. For the background estimate, 1.0\,mBq/kg $^{228}$Th and $^{226}$Ra activities were assumed.


Typically, the thickness of the metalization on HPGe detectors is around 300\,nm. 
Thus for a typical detector with 7.0\,cm height and 7.5\,cm diameter, roughly 
\begin{equation}
7\,\mathrm{cm}\cdot\,\pi\,\cdot\,7.5\,\mathrm{cm}\,\cdot\,300\,\mathrm{nm}\,\cdot\,2.7\,\frac{\mathrm{g}}{\mathrm{cm^3}}\,=\,13.4\,\mathrm{mg} 
\end{equation}
of aluminum are deposited on the detector surface.
Taking the production rates expected from the neutron flux from \cite{cite:ziegler}, roughly 
159 decays/g$\,\cdot$\,0.0134\,g\,$\approx$\,2.1 $^{26}$Al decays per detector per year and 28 decays/g\,$\cdot$ 0.0134\,g\,$\approx$\,0.4 $^{22}$Na decays per detector per year are expected.

A Monte Carlo simulation of 25.2$\cdot 10^{6}$\,\,\,$^{22}$Na, $^{26}$Al and $^{228}$Th decays each,
was performed. 
The MaGe \cite{cite:MaGe} framework based on GEANT4 \cite{cite:geant4} was used
to simulate an array of true coaxial HPGe detectors corresponding to a nominal phase II arrangement of GERDA \cite{cite:gerda}.
The array consisted of seven strings with three detectors each, with the strings being aligned at the same height. The configuration had hexagonal closest packing with a closest radial distance of 15\,mm between the detector surfaces. The distance in height between the detector boundaries was 60\,mm. All detectors were true coaxial n-type 18-fold segmented with 75\,mm diameter and 70\,mm height and a bore hole diameter of 10\,mm \cite{cite:characterization}. 
The  $^{22}$Na, $^{26}$Al, $^{226}$Ra and $^{228}$Th contaminations were simulated as being randomly distributed on the metalized surfaces of all detectors.

\begin{figure}
\begin{center}
\epsfig{file=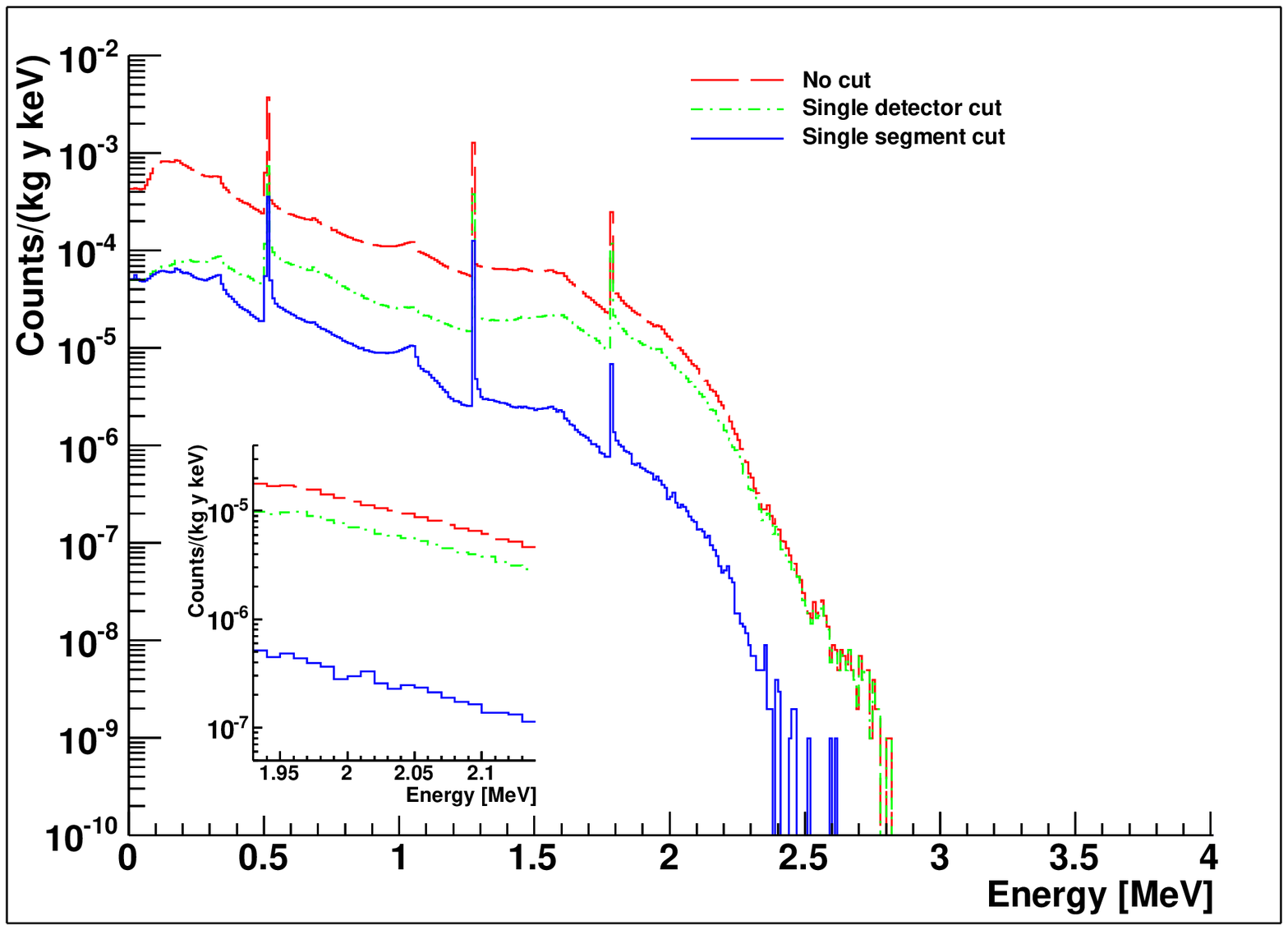, height=12cm}

\end{center}
\caption{Simulated spectrum from a 1.0\,mBq/kg $^{22}$Na contamination in the 
metalization of HPGe detectors in a GERDA like setup \cite{cite:gerda}.
Spectra are without any cut, with single detector cut and with single segment 
cut. The region of interest is shown in the inset.
\label{fig:simulated_22Na}
}
\end{figure}

\begin{figure}
\begin{center}
\epsfig{file=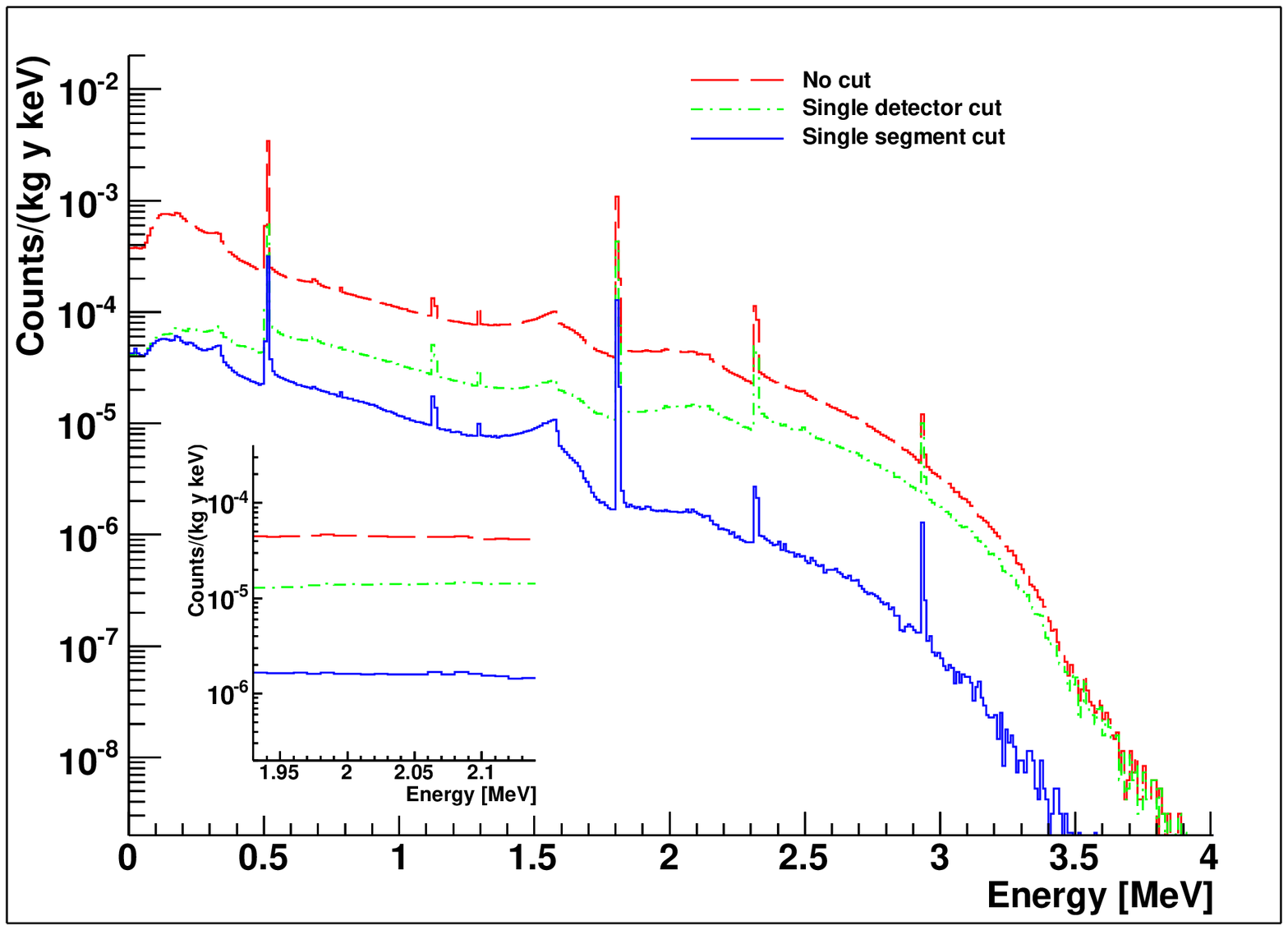, height=12cm}
\end{center}
\caption{Simulated spectrum from a 1.0\,mBq/kg $^{26}$Al contamination in the metalization of HPGe detectors in a GERDA like setup \cite{cite:gerda}. Spectra are without any cut, with single detector cut and with single segment cut. The region of interest is shown in the inset. \label{fig:simulated_26Al}
}
\end{figure}

\begin{figure}
\begin{center}
\epsfig{file=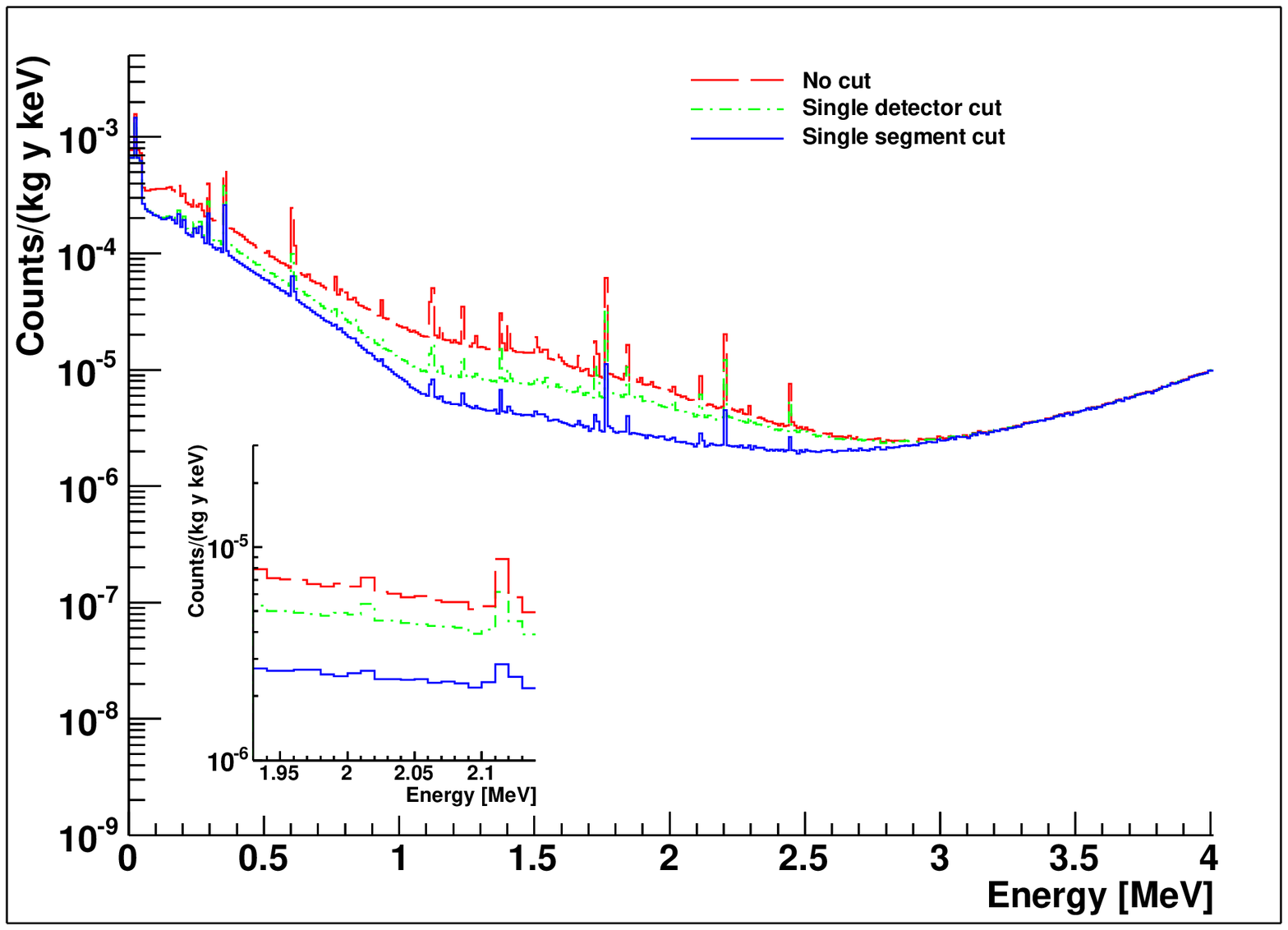, height=12cm}
\end{center}
\caption{Simulated spectrum from a 1.0\,mBq/kg $^{226}$Ra contamination in the metalization of HPGe detectors in a GERDA like setup \cite{cite:gerda}. Spectra are without any cut, with single detector cut and with single segment cut. The region of interest is shown in the inset. \label{fig:simulated_226Ra}
}
\end{figure}

\begin{figure}
\begin{center}
\epsfig{file=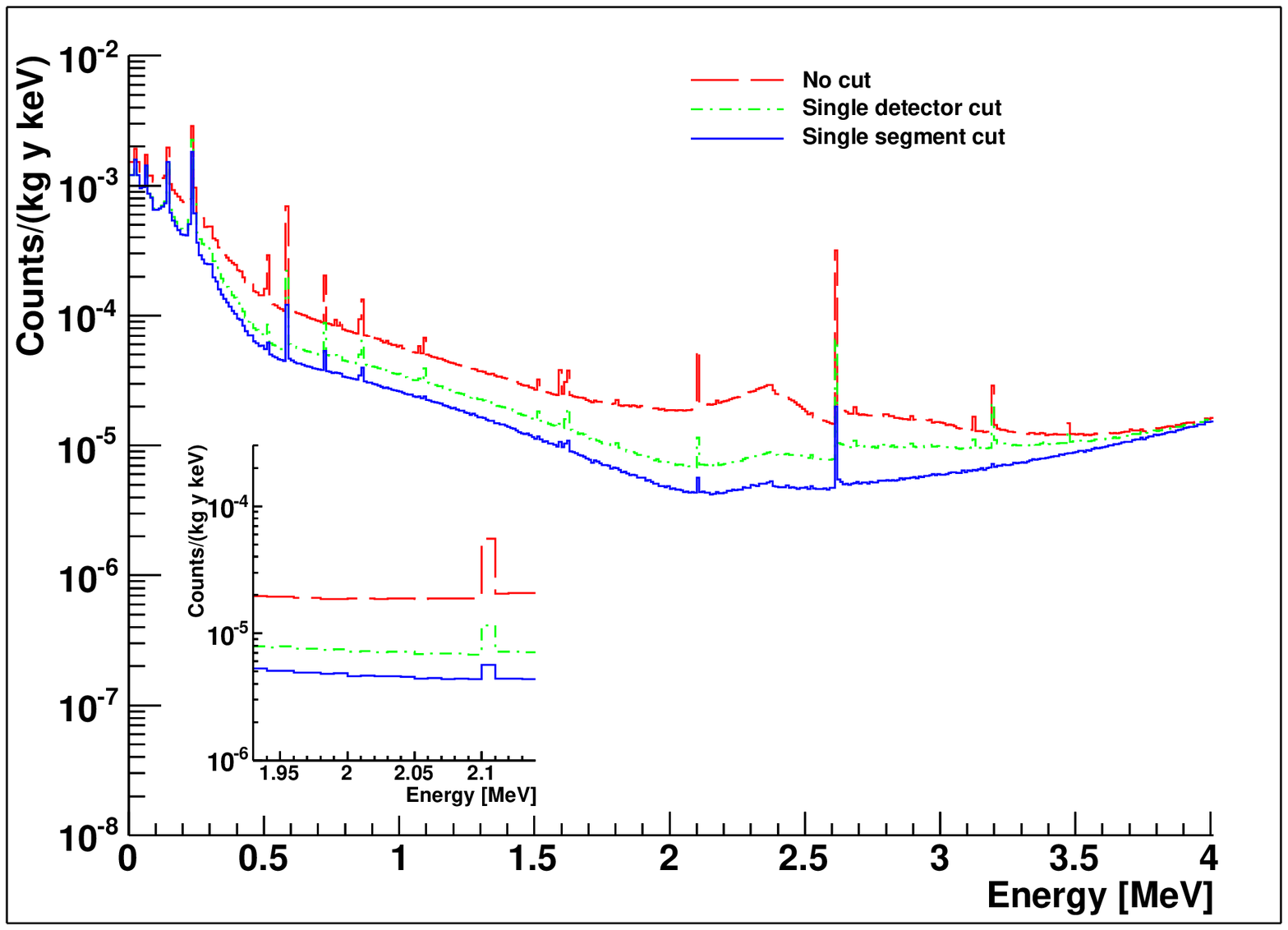, height=12cm}
\end{center}
\caption{Simulated spectrum from a 1.0\,mBq/kg $^{228}$Th contamination in the metalization of HPGe detectors in a GERDA like setup \cite{cite:gerda}. Spectra are without any cut, with single detector cut and with single segment cut. The region of interest is shown in the inset. \label{fig:simulated_232Th}
}
\end{figure}

The results of the simulation are listed in Table 
\ref{tab:sim_background_rates}.
Background events induced by photon interactions can be identified 
through the
event topologies by requiring the energy to be deposited in a single detector or in case of segmented detectors a single segment 
\cite{cite:photon_reduction}. Applying single segment cuts compared to single detector cuts can reduce the background in the region of interest by typically an order of magnitude, depending on the background source. 
The simulated spectra, normalized to the expected number of decays per year and kilogram of germanium, without any cut, with single detector cut and with single segment cut are shown in Figs.\, \ref{fig:simulated_22Na}, 
\ref{fig:simulated_26Al}, \ref{fig:simulated_226Ra} and 
\ref{fig:simulated_232Th}.
The last row of Table \ref{tab:sim_background_rates} gives the tolerable activity of the considered nuclide to restrict the background to 10$^{-6}$ counts/(kg\,y\,keV) with single segment cut applied. 
While for $^{22}$Na an activity of a few mBq/kg can be tolerated, the 
restrictions on $^{26}$Al, $^{226}$Ra and $^{228}$Th contaminations are more sever at the hundreds of $\mu$Bq/kg level.

The resulting spectra in the energy window between 1940~keV and 2140~keV are shown in the insets of Figs.\,\ref{fig:simulated_22Na},
\ref{fig:simulated_26Al}, \ref{fig:simulated_226Ra}  and \ref{fig:simulated_232Th}.
 For $^{26}$Al, the background rate in the region of interest is approximately constant. A background rate of up to 
0.17\,$\cdot\,10^{-5}$\,counts/(kg\,y\,keV) is expected for a single segment cut. 
The background expected from $^{22}$Na is $\approx$\,2.6\,$\cdot$\,10$^{-7}$ counts/(kg\,y\,keV) and thus less 
critical.
The contribution due to a 1.0\,mBq/kg $^{228}$Th contamination is 
expected to be 0.5\,$\cdot\,10^{-5}$\,counts/(kg\,y\,keV),
while for $^{226}$Ra it is 0.26\,$\cdot\,10^{-5}$ counts/(kg\,y\,keV).

If no cut based on event topologies is made, the background rate is higher by roughly one order of magnitude. In this case the background will be at 
levels of up to $10^{-4}$\,counts/(kg\,y\,keV).
A single detector cut reduces the backgrounds due to the decays of $^{22}$Na and $^{26}$Al by factors of 1.4 and 3.2, respectively. 
A single segment gives suppression factors of 10 and 30 for $^{22}$Na and $^{26}$Al, respectively.
For $^{226}$Ra and $^{228}$Th, the single segment cut gives only suppression 
factors of 2.6\, and\,4.5, respectively.

\section{Measurements of $^{26}$Al, $^{22}$Na and primordial nuclides in high purity aluminum samples}
Measurements of five different ULB aluminum samples were performed. All measurements were done to qualify low background aluminum for its use in low background detectors.
The measurements of the samples were carried out at the Laboratori Nazionali del Gran Sasso (LNGS) using the GEMPI \cite{cite:gempi} and GeCRIS \cite{cite:gecris} detectors.
Details of the samples and measured activities or limits are listed in Table \ref{tab:measured_rates}. Some of these results were published earlier \cite{cite:loaiza, cite:koehler}. 
Three of the five samples show a non zero $^{26}$Al activity. The activities 
lie in the range of 0.2\,--\,0.6\,mBq/kg. This translates into 6.3\,--\,18.9 $^{26}$Al decays per year per gram aluminum.

The highest measured activity of 0.6\,mBq/kg of the Kryal \#1 sample, 
see Table \ref{tab:measured_rates}, is close to the one calculated for a 2\,m bauxite layer without top soil. However, for the Kryal \#2 sample, an upper limit of $<$0.26\,mBq/kg was established. 
It is remarkable that the $^{26}$Al activities of the five samples differ so significantly. This indicates that the bauxite used for the production 
of the ULB aluminum was mined at sites with different top soil overburden or with different bauxite layer thicknesses or histories.
It also shows that some of the commercially available bauxite must have originated from mines with no top soil overburden and shallow bauxite layers.

One of the samples shows a $^{22}$Na activity of 0.7$\pm$0.3\,mBq/kg corresponding to 22$\pm$9\,\,\,$^{22}$Na decays per gram aluminum per year. 
This is in very good agreement with the assumption of exposure of the ULB aluminum to sea level secondary cosmic rays for about two years before the screening measurement. Alternatively the sample could have been exposed to increased cosmic radiation by air transport.

In some of the ULB aluminum samples, also $^{226}$Ra or  $^{228}$Th activities were measured. In all cases the $^{228}$Ra activity was considerably lower than the $^{228}$Th activity, suggesting that the $^{232}$Th chain was not in secular equilibrium.
It is not clear, whether $^{228}$Th (and possibly $^{232}$Th) stems from initial contaminations of the bauxite or from impurities introduced to the ULB aluminum during the refining process.


\section{Conclusion}

The detailed geological history of a given mining site 
determines the exposure of the bauxite to neutrons and protons.
Therefore, the source of the aluminum has to be controlled and the aluminum used
has to be screened for radio-impurities with sufficient sensitivity.

The contamination with $^{26}$Al can be avoided by using aluminum refined from
underground bauxite deposits. The contamination with $^{22}$Na has to be 
avoided
by restricting the time of the aluminum stored above ground.
The calculations show that the background induced by  $^{26}$Al, $^{226}$Ra, $^{228}$Th and to a lesser extent $^{22}$Na in the metalization can be significant, if proper care is not taken.

$^{26}$Al has been measured in some ULB aluminum samples 
with activities consistent with the prediction of the calculations taking into account a 2\,m thick bauxite layer without or with very little top soil.
In one of the ULB aluminum samples $^{22}$Na was found
with an activity consistent with two years of sea level exposure 
to cosmic rays.

The simulations show that for neutrinoless double beta decay experiments using HPGe detectors with metalization, the material used to metallize the surfaces has to be carefully screened for radio-impurities. For the case of n-type HPGe detectors with a diameter of 7.5\,cm and a height of 7.0\,cm, maximal radio-impurities of less than 0.6\,mBq/kg, 3.3\,mBq/kg, 0.2\,mBq/kg and 
0.2\,mBq/kg are required in order not to exceed individual background contributions of larger than 10$^{-6}$\,counts/(kg\,y\,keV) from  $^{22}$Na, $^{26}$Al, $^{226}$Ra and $^{228}$Th, respectively. ULB HPGe detectors with the sensitivity required to test ULB aluminum with this sensitivity are available \cite{cite:gempi, cite:gecris}.

%
In order to avoid dangerous radio-impurities, the 
aluminum for the next generation neutrinoless double beta-decay experiments has to be produced from bauxite from mines with sufficient top soil 
overburden. Otherwise, the cosmogenically produced isotope $^{26}$Al will create 
a background level limiting the sensitivity of the experiment.

\section{Acknowledgment}
We would like to thank G. Heusser for the careful reading of the manuscript, many valuable comments and the information about the availability of the HPGe screening measurements of ULB aluminum.

\newpage

\begin{table}
\begin{center}
\begin{tabular}{l|cc|cc}
      &  \multicolumn{2}{c|}{$^{26}$Al} & \multicolumn{2}{c}{$^{22}$Na}\\
      & [(g y)$^{-1}$] & [mBq/kg] & [(g y)$^{-1}$] & [mBq/kg] \\
\hline
Neutrons at sea level& 142 & 4.5 & 56& 1.8\\
Neutrons \cite{cite:neutrons_gordon} & 80 &2.5 & 43&1.3\\
Neutrons in 2\,m bauxite & 21 & 0.67 & 11& 0.4\\
Neutrons, 1\,m quarz sand + 2\,m bauxite & 1.4 &0.04& 1.0 & 0.03\\
Protons at sea level & 17 & 0.54& 8.7 & 0.10\\
Protons in 2 m bauxite & 2.1 & 0.07 & 1.0 & 0.03 \\
Protons, 1\,m quarz sand + 2\,m bauxite& 0.1 & 0.0 & 0.01 & 0.0\\
\hline
\textbf{n+p, 2m bauxite} & 23 & \textbf{0.74}\\
\textbf{sea level, half equilibrium} & & & 32 & \textbf{1.0}\\
\end{tabular}
\end{center}
\caption{Calculated production rates and corresponding activities for secular equilibrium  for $^{26}$Al and $^{22}$Na for different overburden scenarios: at sea level without any overburden, for a 2\,m thick bauxite layer accounting for self absorption and for a 2\,m thick bauxite layer underneath 1\,m top soil.
Secondary neutron spectra from \cite{cite:ziegler} were assumed if not otherwise stated. The last two lines give a realistic prediction for $^{26}$Al, where equilibrium is assumed after millions of years in a 2\,m bauxite layer, and for $^{22}$Na, where half equilibrium is reached at the surface after refinement of the aluminum.
\label{tab:prod_rates}
}
\end{table}

\begin{table}
\begin{center}
\begin{tabular}{l|crr|rcc|c}
Nuclide  & \multicolumn{3}{|c|}{Number of events in RoI} & \multicolumn{3}{|c|}{Background index } & Allowed\\
         & total & SD cut & SS cut              & total & SD cut & SS cut & activity\\
        &      &        &                     & \multicolumn{3}{|c|}{[10$^{-5}$ Counts/(kg y keV)]} & [mBq/kg]\\
\hline
$^{22}$Na & 209\,374 & 122\,942 & 5\,335 &  0.26   & 0.16 &  0.03 & \textbf{3.3}\\
$^{26}$Al & 843\,308 & 26\,828 &  30\,670 & 4.62 & 1.45 & 0.17 & \textbf{0.6}\\ 
$^{228}$Th& 419\,903 & 149\,115 &  93\,647& 2.19 &  0.78 & 0.49 & \textbf{0.2}\\
$^{226}$Ra&   127\,017&93\,907&49\,846 & 0.67 & 0.49  & 0.26 &\textbf{0.4}\\

\hline
\end{tabular}
\end{center}
\caption{The first three columns show the simulated number of events which have an integrated energy deposit between 1940\,keV and 2140\,keV in the detector array (total), in a single detector (SD cut) and in a single segment (SS cut) of a detector. The resulting background indices expected for an activity of 1.0\,mBq/kg in this energy window are given in the next three columns. The last column lists the activity allowed to restrict the background to 10$^{-6}$ counts/(kg\,y\,keV).\label{tab:sim_background_rates}}
\end{table}

{\tiny
\begin{table}
\begin{center}
\begin{tabular}{l|cccccc}
Sample  &$^{26}$Al & $^{22}$Na & $^{226}$Ra & $^{228}$Ra & $^{228}$Th & $^{40}$K\\
\hline
Kryal, Hydro Al., UTH\,1&  \textbf{0.6$\pm$0.3}	&  \textbf{0.7$\pm$0.3} &   $<$ 0.38 &  $<$1.9 &$<$1.7 & $<$21\\
Kryal, VAW, UTH\,0.25  &   $<$ 0.26 &   $<$ 0.15 &  $<$0.58 &  $<$1.2 & $<$0.65 & $<$22\\
Highpural, VAW&  $<$ 0.45&  $<$ 0.37  &   $<$3.7 & \textbf{12$\pm$2}  & \textbf{47$\pm$5}  & $<$5.5\\
ULB I \cite{cite:koehler}&   \textbf{{0.2$\pm$0.1}}	&   $<$ 0.32 & $<$0.7 & $<$0.9 &  \textbf{3.8$\pm$0.7} &  \textbf{4.9 $\pm$1.8}\\
ULB II \cite{cite:loaiza}&  \textbf{0.38}$^{ \textbf{+0.19}}_{\textbf{\,--0.14}}$ &  $<$ 0.18 &  \textbf{0.27$\pm$0.19} &   $<$ 0.11 & \textbf{1.4$\pm${0.2}} &  \textbf{1.1}$^{ \textbf{+0.2}}_{ \textbf{\,--0.1}}$\\
\hline
\end{tabular}
\end{center}
\caption{Measured $^{26}$Al, $^{22}$Na, $^{226}$Ra , $^{228}$Th and $^{40}$K activities of  ULB aluminum samples in mBq/kg. \label{tab:measured_rates}
}
\end{table}
}

\newpage

\end{document}